\documentclass{article}

\usepackage{times}
\usepackage{graphicx} 
\usepackage{subfigure} 

\usepackage{natbib}

\usepackage{algorithm}
\usepackage{algorithmic}

\usepackage{hyperref}

\usepackage[accepted]{icml2017}

\icmltitlerunning{Beyond the technical challenges for deploying Machine Learning solutions in a software company}

\begin{document} 

\twocolumn[
\icmltitle{Beyond the technical challenges for deploying Machine Learning\\ solutions in a software company}




\begin{icmlauthorlist}
\icmlauthor{Ilias Flaounas}{atlassian}
\end{icmlauthorlist}

\icmlaffiliation{atlassian}{Atlassian, Australia}

\icmlcorrespondingauthor{Ilias Flaounas}{iflaounas@atlassian.com}

\icmlkeywords{machine learning, ICML}

\vskip 0.3in
]



\printAffiliationsAndNotice{} 

\begin{abstract} 
Recently software development companies started to embrace Machine Learning (ML) techniques for introducing a series of advanced functionality in their products such as personalisation of the user experience, improved search, content recommendation and automation. The technical challenges for tackling these problems are heavily researched in literature. A less studied area is a pragmatic approach to the role of humans in a complex modern industrial environment where ML based systems are developed. Key stakeholders affect the system from inception and up to operation and maintenance. Product managers want to embed ``smart''  experiences for their users and drive the decisions on what should be built next; software engineers are challenged to build or utilise ML software tools that require skills that are well outside of their comfort zone; legal and risk departments may influence design choices and data access; operations teams are requested to maintain ML systems which are non-stationary in their nature and change behaviour over time; and finally ML practitioners should communicate with all these stakeholders to successfully build a reliable system. This paper discusses some of the challenges we faced in Atlassian as we started investing more in the ML space. 
\end{abstract} 

\section{Introduction}

\begin{table*}[t]
\caption{Key stakeholders and challenges for each phase of a building a ML based product feature}
\label{sample-table}
\vskip 0.15in
\begin{center}
\begin{small}
\begin{tabular}{lcccr}
\hline
\abovespace\belowspace
 & Ideation & Execution & Operation \\
\hline
\abovespace
Key   & Product Managers & Data Scientists & SRE \\
Stakeholders	& Designers & Data Engineers & Support \\		
 &  &  Developers & \\
 \\
\hline
\\
 &  Data availability &  Build vs rent & Maintain accuracy\\
 Challenges &  Privacy concerns &  Scalability issues & Stability of data sources \\
 \belowspace
	  & Project risk estimation & Productionisation & \\

\hline
\end{tabular}
\end{small}
\end{center}
\vskip -0.1in
\end{table*}

In the last few years, software development companies have changed from building and ``shipping'' their products to customers to become online service providers and offer Software as a Service (SaaS). Their products live in the Cloud and customers connect to them online. This change offers software companies some new benefits such as the access to detailed behavioural data of their users. Storage and computation costs have dropped significantly allowing the analysis of these vast data sources and leading to data-driven decision making processes that further improves the products. 

Recently, software companies started to experience the next big paradigm shift in their operation. The bar for utilising ML techniques has been lowered significantly. Even advanced modelling techniques have started to become a commodity. As discussed companies already hold a plethora of data about their users which can be used as training data for ML algorithms. Furthermore, because products live in the cloud it is feasible to constantly improve their products on the fly and thus increasing the appetite for experimentation. However, we are also experiencing a lot of hype driven mainly by the media, consultants and vendors. This becomes more challenging given that companies typically lack the scarce relevant talent required to drive them successfully in this new era. 

In a more pragmatic approach Data Scientists, defined here as ML practitioners, unfortunately are a scarce resource absorbed by the few big orgs that pioneer the field. Thus non-experts are getting involved in the development of the ML system and this creates a less compelling environment in most cases in the rest of the industry. It is a learning process for most parties involved in the process.

Table I summarises the key stakeholders and the challenges they face for the different phases of the development of a ML based product feature. The challenges listed here are those beyond the typical expected daily tasks of ML experts such as cleaning data, preprocessing, training and evaluating models until some acceptable level of performance is achieved. Note that the proposed bucketing of roles and challenges is not absolute: different companies --- and even different projects within the same company --- may encounter a different set of challenges or experience an active involvement of key stakeholders across multiple phases of the project.

ML based projects can be divided in three phases: Ideation, Execution and Operation. In the ideation phase we explore the potential impact of an ML system and ultimately make a decision on whether the idea deserves funding and the allocation of resources (both human and computational). In the execution phase data-scientists will come up with some early exploration of the potential in the form of a prototype. An experiment could provide some quantitative validation of the system. If all goes well then a development team needs to polish and ``productionise'' the prototype, while the data engineers team needs to secure the stability and availability of the required data sources. Practice shows that the productionasition of prototypes may seem like the last 10\% of the project but can actually take 90\% of the time without some provisions. After the feature is complete it is surfaced to end-users. After that point typically Site Reliability Engineering (SRE) team will monitor its performance. 

As the table suggests there is a large range of different stakeholders involved in developing and operating an ML system. The human factor becomes a critical component for success. Some of the challenges include: stakeholders are not aware of the effort needed to build such a system; which are the hard parts and which are the easy parts? How difficult is that last 5\% that is missing to bring the system in the hands of real users? 

This paper discusses some of the challenges we faced in Atlassian, the role of the different stakeholders and the challenges we encountered as we started investing more in ML based solutions.

\section{Ideation phase: impact and feasibility of the system}
In the ideation phase, a product team starts to consider an idea for a new or improved product functionality that requires some notion of automation or personalisation. Typically simple heuristics should be considered first as a potential solution. If a more accurate solution is required, or the heuristics are too complex then this could be an excellent candidate for an ML based solution.
Product managers need to decide whether the idea should get the required funding. To make that decision two main factors need to be considered: the size of the opportunity and the feasibility of the idea. 

The size of the opportunity is an estimation of how many users will get value from the new system. Is it  the entire user base or a small segment of it? For example, a content recommendation system provides value only if the volume of content is large enough so that simple browsing is not an adequate solution or if the pace of creation of new content would require the constant attention of users. Product managers would need some help to evaluate the magnitude of the impact that such a system would have on their users. This should be compared to the value provided from existing functionality. In the case of the content recommendation system, that would be the current browsing or search capabilities of the product. In summary, the new ML system should add value on top of that provided by the current functionality and provide that value to an adequate number of end users. 

Another factor to be considered is the feasibility of the idea, given limited resources in time, people and data. This requires some feedback from ML practitioners. The questions to be answered include: what is the right metric we try to optimise for? Are the data and infrastructure required for operation currently available? How confident are we that a ML system can be built with some minimum level of performance in key metrics like mean square error, or precision and recall? Currently there is not much research that helps practitioners estimate the anticipated performance of the system given the data available, in terms of samples, features and some methodology. They need to rely on rules of thumb, their experience and previous efforts in similar systems to make an estimation. Some follow up questions include how easy it would be to maintain such a system, and whether there are any privacy concerns. While answering all these questions requires subject matter expertise, they are outside the typical training of a ML practitioner on building a learning system.

A very sensitive area that affects decisions on the feasibility or the design of a ML system are concerns around the privacy of users and their data. The legal framework within which ML systems operate is changing fast as governments try to regulate the access of companies to the private data of their citizens. Legal and risk departments of companies are getting involved and the design of ML based systems is severely affected in an effort to provide compliance with the privacy policy of the organisation and the legal framework. A new European regulation, namely the ``General Data Protection Regulation'' (GDPR) that will take effect in May 2018, introduces new requirements for companies and strict rules on how user data can be used \cite{goodman2016european}. There are still quite a few unknowns and blurry areas, however the new challenges for ML practitioners are to balance company policies and procedures, comply with the new upcoming legislation, respect of user privacy, and at the same time design automation and product functionality that requires access to the raw data. Some hard questions include how to debug a ML system at scale when there is no access to the raw data generated by users? Can users be profiled based on their sex, age, experience, their role? If not, can these attributes indirectly be inferred from other known variables and fed into a ML system? The building of a ML system is a challenging experience considering that such systems are the central places where data from different data sources and potentially different users are integrated, in non-interpretable ways, to offer the desired functionality.

\section{Execution phase: the building of the system}

Obviously ML practitioners are the most qualified people to create ML solutions. However, their skills are a scarce resource for most development teams. A typical engineering team does not have the experience to deal with this set of problems \cite{Zinkevich}. Some solutions include the renting of external services or the use of consultants. However, external consultants can not support and maintain the system after its delivery, as discussed further in the next section. 

In our days, there are a lot of online service providers and off-the-self-tools that promise accessible ML solutions for any type of problem. These tools and services combined with hype over technological breakthroughs and a lack of domain expertise create a problematic environment. Non-experts can easily believe that a plug-n-play deployment of the latest ML tool will solve the problems at hand. The expectations are raised and morale is degraded if the prototypes they build underperform. Indeed, picking the correct ML tool or algorithm is just a small part of the requirements for building a full ML system. Many times the hard part is training and tuning the algorithm; providing the correct inputs, cleaning up and preprocessing the data. Sometimes a solution may require further post processing and ad-hoc rules for edge cases. 

When a prototype is ready it will be deployed to some end users. Only then some problems will be revealed. For example, maybe the system does not achieve the minimum level of required performance; or maybe the target concept is not learnable for all users and all the different ways they use the product; or maybe more data is required. Problems like these may be hard to detect when building a prototype and before deploying to real users. While this risk is well understood by ML practitioners and they can provision some work arounds, it is a quite a novel experience for software engineers that are accustomed to working with deterministic systems.

Another class of common problems is around the scalability challenges of traditional ML techniques. While an individual ML system may be reasonably fast to train and in making predictions, it may actually be quite slow overall when we consider the number of times it needs to be deployed. Suddenly even a few-minutes training time becomes a blocker if the training must be repeated for millions of users. Techniques that assume unbounded resources, for example the construction of a ``full'' dictionary of term frequencies, or techniques that scale linearly with the number of samples suddenly appear as quite slow performers.

Even simpler problems that require the creation of back-end business services suffer from unique challenges. For example, in our company there was a need to automate the monitoring of some business metrics and provide automated alerts for anomalies in the data. The development team tasked to solve the issue explored the options to either rent or build a relevant service. Both choices had a caveat though: there should be some human feedback to tune and maintain the system, specifically a process to notify the system of false positives and false negatives. From the development team's perspective that looked like a minor step in the process, compared to the challenge of creating the algorithm that detects the anomalies. On the other hand, the data scientists that were pulled in for domain expertise all reached the consensus that building such a system would be easy, if not trivial, but the tuning phase would be the real challenge and the time consuming step. They advised that the system will require full re-training, for example in cases of data loss/delays, or annotation of data points that represent one-off well understood events, and in cases of more permanent changes a mechanism for the system to become less sensitive until a new baseline is established.

\section{Operation phase: the maintenance of the system}

The maintenance and operation of a ML based system is a significant challenge. The system faces not only the challenges of any other software system, but even more that relate to its non-deterministic nature that changes depending on the data input. If the system is poorly designed, then its operation is potentially harder than the actual building of the system. 
An implicit assumption when building an ML system is that all the underlying data sources will be available, stable and with similar statistics as those encountered in the training phase. Even if one data source becomes unavailable the whole system's performance may degrade to zero. In practice, the assumption of a stationary environment is problematic, especially when the system is designed to consume data sources that are typically outside of the control of the team that built it. While to some extent the missing data may be considered as input noise from a system designer's perspective, in practice this ``noise'' may sustain for a significant amount of time, saturating online learning systems. The ML practitioners end up trading off the performance of the system for stability and reliability \cite{sculley2015}. 

In practice, the operation of an ML system requires constant monitoring of some key performance metrics. Monitoring the system performance requires also training development operations on what signals need to be monitored and what needs to be done when things go wrong. Some best practices for monitoring such systems are discussed in \cite{breck2016}. 
Sometimes the system performance may degrade to a level that makes it unusable. A solution in those cases would be to use feature flags, a software development methodology that allows turning off or swapping backend services on the fly. This way the system may fail over to some basic functionality until the problem is solved.

A relevant policy typically found in a software company is summarised in the phrase ``you built it, you maintain it''. This policy renders development teams responsible for the proper operational status of the services and systems they build. They can not build something suboptimal and expect the operations and support teams to maintain it. In the case of the ML systems that policy means that the data scientists need to create stable systems with provisions for missing or noisy data sources. Otherwise they will end up continuously maintaining and tackling problems one after the other, focusing all their time in one single system. The problem becomes worst if ``edge cases'' need to be tackled separately and are ignored at design. In a system that is used by millions of users the edge cases are actually expected to be encountered quite often.

\section{Conclusions}

The development of an operational ML system requires the involvement of many different stakeholders. These people may be experts in their respective fields and they all need to cooperate to successfully productionise a ML system. The ML practitioner needs to provide input for a series of decisions outside her strict domain of expertise and beyond the typical task of training a ML system. The development of the system requires taking into account factors that balance the performance with the scalability and the cost of long term maintenance of the system.

\section*{Acknowledgements} 
Author is grateful to Arik Friedman, Daniel Louw and Hercules Konstantopoulos for discussions and feedback; and the anonymous reviewers for their constructive comments.

\bibliography{my_paper}
\bibliographystyle{icml2017}

\end{document}